\documentclass[twocolumn]{aastex61}  

\def\deg{$^{\circ}$~}
\def\phr{~ph~cm$^{-2}$~s$^{-1}$~}
\def\unorm{~TeV$^{-1}$~cm$^{-2}$~s$^{-1}$~}

\begin{document}


\title{Daily monitoring of TeV gamma-ray emission from Mrk 421, 
Mrk 501, and the Crab Nebula with HAWC}

\AuthorCallLimit=150

\correspondingauthor{Robert J. Lauer}
\email{rjlauer@unm.edu}

\author{A.U.~Abeysekara}
\affil{Department of Physics and Astronomy, University of Utah, Salt Lake City, 
UT, USA } 

\author{A.~Albert}
\affil{Physics Division, Los Alamos National Laboratory, Los Alamos, NM, USA }

\author{R.~Alfaro}
\affil{Instituto de F\'{i}sica, Universidad Nacional Aut\'{o}noma de M\'{e}xico, 
Mexico City, Mexico }

\author{C.~Alvarez}
\affil{Universidad Aut\'{o}noma de Chiapas, Tuxtla Guti\'{e}rrez, Chiapas, 
M\'{e}xico}

\author{J.D.~\'{A}lvarez}
\affil{Universidad Michoacana de San Nicol\'{a}s de Hidalgo, Morelia, Mexico }

\author{R.~Arceo}
\affil{Universidad Aut\'{o}noma de Chiapas, Tuxtla Guti\'{e}rrez, Chiapas, 
M\'{e}xico}

\author{J.C.~Arteaga-Vel\'{a}zquez}
\affil{Universidad Michoacana de San Nicol\'{a}s de Hidalgo, Morelia, Mexico }

\author{D.~Avila Rojas}
\affil{Instituto de F\'{i}sica, Universidad Nacional Aut\'{o}noma de M\'{e}xico, 
Mexico City, Mexico }

\author{H.A.~Ayala Solares}
\affil{Department of Physics, Michigan Technological University, Houghton, MI, 
USA }

\author{A.S.~Barber}
\affil{Department of Physics and Astronomy, University of Utah, Salt Lake City, 
UT, USA }

% ...



\begin{abstract}
We present results from daily monitoring of gamma rays in the energy 
range $\sim0.5$ to $\sim100$~TeV with the first 17 months of data from 
the High Altitude Water Cherenkov (HAWC) Observatory.
Its wide field of view of 2 steradians and duty cycle 
of $>95$\% are unique features compared to other TeV observatories 
that allow us to observe every source that transits over HAWC 
for up to 
$\sim6$~hours each sidereal day.
This regular sampling yields unprecedented light curves from unbiased 
measurements that 
are independent of seasons or weather conditions.
For the Crab Nebula as a reference source we find no variability in the TeV 
band. 
Our main focus is the study of the TeV blazars Markarian (Mrk) 
421 and Mrk 501.
A spectral fit for Mrk 421 yields a power law index $\Gamma=2.21 
\pm0.14_{\mathrm{stat}}\pm0.20_{\mathrm{sys}} $ and an exponential cut-off 
$E_0=5.4 \pm 1.1_{\mathrm{stat}}\pm 1.0_{\mathrm{sys}}$~TeV. 
For Mrk 501, we find an
index $\Gamma=1.60\pm 0.30_{\mathrm{stat}} \pm 0.20_{\mathrm{sys}}$ and 
exponential cut-off $E_0=5.7\pm 1.6_{\mathrm{stat}} \pm 
1.0_{\mathrm{sys}}$~TeV.
The light curves for both sources show clear variability and a Bayesian 
analysis is applied to identify changes between flux 
states. The highest per-transit fluxes observed from Mrk 421 exceed 
the Crab Nebula flux by a factor of approximately five. For Mrk 501, 
several transits show fluxes in excess of three times the Crab Nebula flux.
In a comparison to lower energy gamma-ray 
and X-ray monitoring data with comparable sampling we cannot 
identify clear counterparts for the most significant 
flaring features observed by HAWC.

\end{abstract}


\section{Introduction}

Most extragalactic sources of gamma-ray emission at TeV 
energies are blazars, active galactic nuclei (AGN)
with jets oriented close to the line of sight.\footnote{See list at 
http://tevcat.uchicago.edu .}
Due to the steep viewing angle and the limited angular resolution of 
TeV observations, the locations of such regions are not resolved in GeV or TeV 
observations.
The general consensus is that a rotating central black hole serves as power
source, transporting energy along the jets to one or multiple emission regions.
In competing model descriptions, the conversion into kinetic energy happens 
either through stochastic acceleration in relativistic shocks or through 
magnetic 
reconnection; see~\citet{Sironi2015} and references therein for a 
recent overview.
Depending on the dominating population of accelerated particles being either 
electrons or protons, models are also 
categorized as either leptonic~\citep{Rees1967} or 
hadronic~\citep{Mannheim1993}.
The latter would provide a framework for AGNs as sources 
of charged cosmic rays and neutrinos. Such 
hadronic models for gamma-ray emission from blazars have recently been found 
to be hard to reconcile with measurements 
of jet power~\citep{Zdziarski2015}, but this might not hold true for high energy
peaked BL Lac objects with a gamma-ray peak in the spectral energy distribution 
around 1 TeV~\citep{Cerruti2014}.
Acceleration models generally have to allow for variability in the TeV 
emission, 
since a number of TeV blazars are known to exhibit strong flux changes. 
During such flares, TeV fluxes have been 
observed to increase by an
order of magnitude and to vary on time scales from 
months down to minutes~\citep[see 
e.g.][]{Aharonian2007PKS2155,Albert2007Mrk501}.
The monitoring of TeV gamma-ray variability can provide critical insights into 
the energetics and mechanisms of acceleration. 
Long-term observations of TeV variability are particularly 
valuable to establish flaring frequencies and variability time scales can be 
used to constrain sizes of emission regions. 
By putting unbiased TeV data in the context of multiwavelength 
observations
we can test if we see strong
correlations across energies as expected in one-zone models or if a 
multi-zone description is required~\citep[see e.g.][]{FermiEraMrk501}.
Furthermore, systematic tests of correlations 
between TeV gamma rays from blazars and multimessenger data, such as IceCube 
neutrino signals, can benefit from regular monitoring.
  
Most of the observations in the TeV band have been performed with imaging 
atmospheric Cherenkov telescopes (IACTs) that can only operate during clear 
nights and 
typically monitor only one source in the field of view at any time. 
Atmospheric conditions and competing observation tasks generally 
limit the time available for long-term studies of individual objects. 
Observations of blazars have also often been biased by the tendency to 
follow up on flare alerts, preventing equal, unbiased coverage of low flux 
states or flares without multiwavelength correlations.
The monitoring program of the First G-APD Cherenkov Telescope 
\citep[FACT;][]{FACT2013} aims at unbiased scheduling of regular 
observations for 
selected objects,
but the observations are still limited by seasonal visibility constraints and 
weather.
Current IACTs can therefore provide very useful data for in-depth studies of 
individual flares but have only limited capabilities for regular, systematic 
monitoring.
Previous long-term blazar monitoring studies by wide field-of-view TeV 
instruments with high duty cycles can be found, for example, 
in~\citet{MilagroMrk4212014,ArgoMrk4212011,ArgoMrk5012012}
but only include light curves that integrated over week- or month-long 
intervals due to limited sensitivity.

With the High Altitude Water Cherenkov (HAWC) Observatory we now 
have a very high energy instrument
that can  
monitor any source over two thirds of the sky 
for up to 6~hours per day.
These capabilities make unprecedented TeV light curve
data available for studying flaring behavior of blazars. 
In addition, scanning a large part of the sky with this sensitivity will 
increase the chances to find bright flare events from established and 
new extragalactic sources that can be used to constrain or measure the 
extragalactic background light~\citep[EBL;][]{Stecker1992} and intergalactic 
magnetic fields~\citep{Neronov2007}. 

Preliminary blazar light curves from data taken in 2013 and 2014 with the 
partial HAWC array were shown in~\citet{blazars-icrc2015}.
In this paper we present the first long-term TeV light curve studies with 
single-transit intervals
that are based on data from the completed HAWC Observatory, taken over 
17 months between 2014 November and 2016 April. We are focusing on 
the two blazars that have been significantly detected in the second 
HAWC catalog~\citep{HawcCatalog2016}, Markarian (Mrk) 421 and Mrk 501.
After a short discussion of the instrument in Section~\ref{sec:det}, we 
describe the analysis methods in Section~\ref{sec:analysis}, including the
production of light curves via daily sky maps, the maximum likelihood analysis 
for deriving flux and spectral measurements and the algorithms for 
characterizing
variability.
To verify the variability analysis on a reference source, we apply these 
methods to 
the Crab Nebula in Section~\ref{sec:crab}. 
We then present the main results from applying the analysis to Mrk 
421 in Section~\ref{sec:mrk421} and Mrk 501 in 
Section~\ref{sec:mrk501}.
A discussion of the results is included at the end of each of the sections 
for the individual sources, and we close with conclusions and outlook in 
Section~\ref{sec:outlook}.

\section{The HAWC Observatory}
\label{sec:det}

The HAWC
Observatory is located at an elevation of 4,100~m above sea level on the 
flanks of the Sierra Negra volcano in the state of Puebla, Mexico 
(97.3$^{\circ}$W, 
19.0$^{\circ}$N). Covering an area of 22,000~m$^2$, the array consists of 300 
water Cherenkov detectors (WCDs), each filled with 190,000 liters of 
water and 
instrumented with 4 photomultiplier tubes (PMTs) to detect Cherenkov light from 
charged particles in extensive air showers. Light-tight bladders inside the 
corrugated steel frame optically isolate each detector from the environment 
which allows HAWC to 
be operated continuously, with down time only due to maintenance.
Before HAWC was completed in 2015 March it had been operating in a partial 
configuration with 250 WCDs since 2014 November, leading to a slight 
improvement during the first few months of data included in this 
paper.

The design of HAWC is optimized for the detection of air showers induced by 
gamma rays between $\sim 0.1$ and $\sim 100$~TeV. Peak sensitivity is reached 
at a few TeV, depending on source spectra. The footprint of an air shower is
recorded through the collection of PMT signals induced by the passing of the 
shower front through the array and is referred to as an event in the 
following. Charge and timing information 
are calibrated via optical laser pulses and are used to 
reconstruct the direction of the primary particle.
Events are sorted into nine analysis bins, defined 
by the fraction of PMTs with signals in a narrow time window (550~ns). Each 
bin has 
individual background suppression cuts that reject a large fraction of showers 
from hadronic primaries based on the distribution of observed charges, which 
includes more bright signals from muons outside the shower core and is less 
smooth for hadronic showers compared to those induced by gamma rays.
The increasing number of PMT signals 
available for direction reconstruction leads to an angular 
resolution\footnote{The angular resolution is defined here as 68\% containment 
radius for events from a point source.} that 
improves from $\sim1$\deg to $\sim0.2$\deg from the first to the last bin.
By quantifying the size of the shower on the ground the bins serve as an energy 
proxy. 
In the analysis presented here, we use the energy distributions in each 
bin predicted by simulation to perform likelihood fits of fluxes and spectra via 
a forward-folding method.
A full description of HAWC data reconstruction, analysis method, performance, 
and systematic uncertainties is 
presented in~\citet{HawcCrab}.

\section{Analysis Methods}
\label{sec:analysis}

\subsection{Sidereal Day Sky Maps}
\label{sec:maps}

\begin{figure}[t]
\centering
\includegraphics[width=0.47\textwidth]{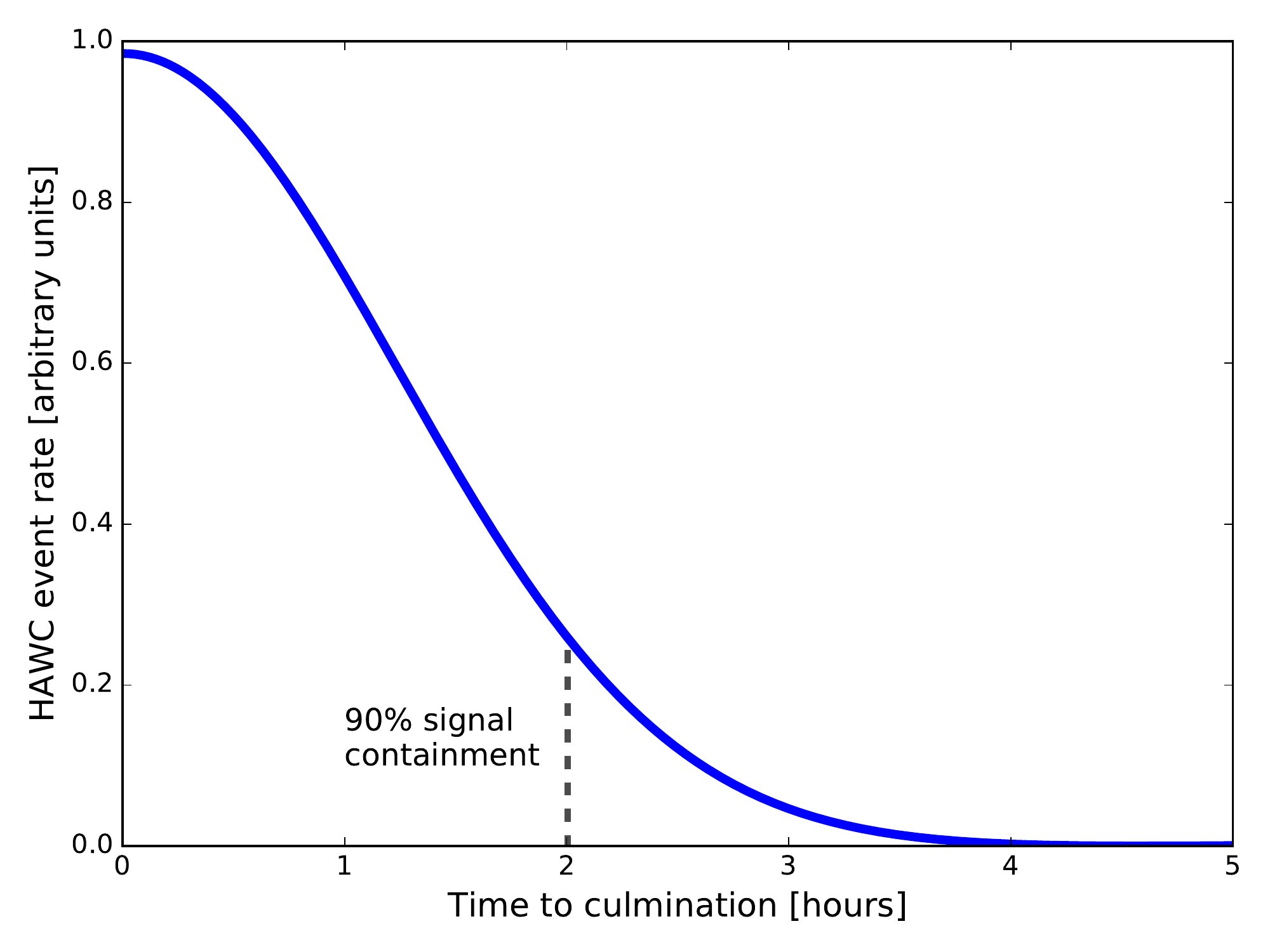}
\caption{Gamma-ray signal rate in HAWC versus time, showing one half of a 
symmetric transit before/after culmination at 0 hours. This distribution is 
based on simulations for a point source at declination 20\deg with a Crab-like 
simple power-law spectrum, photon index $\Gamma=2.63$, and highlights the time 
window during which 90\% of the events are expected.}
\label{fig:zenithcoverage}
\end{figure}

HAWC can record extensive air showers from all directions 
visible above the horizon. Due to 
the increasing absorption of secondary 
particles in the atmosphere, the actual effective area for gamma rays is a 
function of the zenith angle of the primary particle 
and the contribution of events from outside a cone with 
an opening angle of $\sim 45$\deg around zenith is usually small. 
The field of view thus spans a solid angle of 
$\sim2$~steradians (sr) and HAWC is most sensitive to sources 
between declinations $-26$\deg and $+64$\deg. With the rotation 
of the Earth, any location in this declination range passes over HAWC once 
every sidereal day. 
In the following, a transit is defined by visibility over HAWC at zenith 
angles $\theta<45$\deg and lasts approximately 
6 hours for the sources discussed in this paper.
The detection efficiency is not uniform during 
the transit and Fig.~\ref{fig:zenithcoverage} shows the expected fraction of 
signal as a function of time relative to culmination. 
Approximately 90\% of the signal events arrive within the central 
$\sim 4$~hours of a transit for a source modeled on the Crab Nebula (photon 
index $\Gamma=2.63$ and declination 22\deg).
While the shape of this event distribution as a function of transit 
time can in principle change for different spectra and declinations, it is not 
significantly altered for the sources discussed in this paper, which 
culminate within $\leq 20$\deg of zenith. 
For the flux measurement over a full transit, the zenith dependence 
of HAWC's sensitivity simplifies to a dependence on the source's 
declination that determines the expected excess and energy distribution. 

In order to process the data in units that do not contain more than one full 
transit for any source, all reconstructed events are 
sorted into sidereal days, starting at midnight local 
sidereal time at the HAWC site.\footnote{This choice 
leads to 
transits being split in two for sources with right ascension~$<3$~h or 
$>21$~h. Such sources, though not discussed in this paper, can be 
analyzed with a separate set of maps binning the data with their 
start times offset by 12 sidereal hours.}
For each sidereal day and each of the nine analysis bins, a sky map of event 
counts is produced by populating pixels on a HEALPix grid~\citep{HEALPix} 
with 
an average spacing of $\sim0.06^{\circ}$. These maps are still 
dominated by hadronic background events and we use direct 
integration~\citep{Atkins2003} to obtain a background estimate. In this 
procedure, a local efficiency map is created by averaging counts in a strip 
of pixels over two hours in right ascension around any location. We smooth 
this efficiency map via a spline fit to compensate for the limited 
statistics in higher analysis bins. The pixels 
near the strongest known sources and the galactic plane are excluded during 
the 
averaging in order not to bias the result by counting gamma ray events as 
background. Due to the limited statistics in higher analysis bins, we perform a 
spline fit of the local efficiency distributions during the direct integration 
procedure.
The estimated background counts in each pixel are stored in a second map with 
the same grid structure.

A quality selection is applied before including data in 
the maps. 
First, monitoring of the stability of the angular distributions of 
reconstructed background events is used to exclude
data taken during unstable conditions, for example related to 
maintenance. 
In order to control rate stability during a 
sidereal day, we then fit the detector rate with a function that follows 
tidal effects of the atmosphere and reject short 
periods of data that significantly deviate from this fit.
To ensure a uniform detector response, sidereal days with partial 
coverage are not included for a given source if the lost signal fraction is 
expected to exceed 50\%. This expected coverage fraction is calculated 
by integrating the signal distribution from Fig.~\ref{fig:zenithcoverage} only 
over those sections of the transit that are included in the data, assuming a 
uniform flux during 6 hours.
The different right 
ascensions of the three sources lead to slightly different exposures which are 
reported in Table~\ref{tab:coverage}.
The total observation time is 
calculated based on 6 hours of effective HAWC observations for an uninterrupted 
transit and is corrected for gaps in case of partial coverage. For the period 
of 513 sidereal days included in this analysis, on average 
92\% of transits or 22\% of actual time per source are covered. 

\begin{table}[t]
\centering
\caption{Observation time per source after quality cuts}
 \begin{tabular}{ccccc}
\hline
\hline
 Source & Included Transits & Time At Zenith Angles $<45$\deg\\
        &                   & [hours]\\
\hline
Crab      &   472   &   2700
\\
Mrk 421   &   471   &   2665
\\
Mrk 501   &   479   &   2750
\\
\hline
\end{tabular}
 \label{tab:coverage}
\end{table}

\subsection{Flux and Spectral Analysis}
\label{sec:FluxSpectral}

For this light curve analysis, the standard HAWC maximum-likelihood 
method~\citep{liff-icrc2015} is 
applied to the sidereal day maps in order to fit photon fluxes for each 
transit of selected source locations. 
The two extragalactic sources discussed here are modeled as gamma-ray 
point sources with differential flux energy spectra described by a power 
law with normalization $F$ at 1~TeV, photon index $\Gamma$ and an optional 
exponential cut-off $E_0$:
\begin{equation}
\label{eq:spectrum}
 \frac{dN_{\mathrm{ph}}}{dE} = F  
\left(\frac{E}{1 \mathrm{TeV}}\right)^{-\Gamma}  
\exp{\left(-\frac{E}{E_0}\right)}\quad.
\end{equation}
In the HAWC likelihood analysis framework, this input flux is convolved with a 
detector response function that 
includes the point spread function and efficiency of triggers and cuts, 
depending on primary energy and incident angle. For one source transit over 
HAWC, the signal hypothesis contributions as a function of zenith angle are 
summed and yield the expected number of events $S_{b,p}$ per analysis bin $b$ 
(ranging from 1 to 9) and pixel $p$ (for all pixels within a 
radius of 3\deg around the source).

In cases where the coverage of a source transit is interrupted, for example 
due 
to detector down time, the lost signal fraction 
compared to a full transit is calculated by excluding the gap period from the 
integration over zenith angles (see Fig.~\ref{fig:zenithcoverage}) and 
the expected event count is reduced accordingly.
For the source hypothesis defined by $\{F,\Gamma,E_{0}\}$ and the 
observation $\mathbf{N}$ of numbers of events in all bins and pixels, we 
express 
the likelihood as  
\begin{equation}
\label{eq:llh}
 \mathcal{L}_S \left( \mathbf{N}, \{F,\Gamma,E_{0}\}\right) = 
\prod_b{\prod_p{ 
P(N_{b,p},\lambda_{b,p})}} \quad ,
\end{equation}
where $P(N_{b,p},\lambda_{b,p})$ is the Poisson distribution for a mean 
expectation $\lambda_{b,p} = S_{b,p}+B_{b,p}$, the sum of the expected 
signal ($S$) and the number of background ($B$) events estimated from data for 
analysis bin $b$ and pixel $p$.
In the likelihood ratio test, the result of equation~(\ref{eq:llh}) is 
compared to 
the likelihood value $\mathcal{L}_B$ for a background-only assumption ($S_{b,p} 
= 0$). 
We express this ratio as the difference of the logarithms of the two likelihood 
values and define 
the standard test statistic as  
\begin{equation}
 \label{eq:de}
 \mathrm{TS} = 2  \Delta\ln{\mathcal{L}} = 2 \left( \ln( \mathcal{L}_S ) -
\ln( \mathcal{L}_B ) \right) 
\quad .
\end{equation}
TS is then numerically maximized by iteratively changing the input 
parameters, yielding those values that have the highest likelihood of 
describing the observed data for the point source model assumption.

For the analysis in this paper, the normalization $F$, the spectral index 
$\Gamma$, and the 
cut-off value $E_0$ in equation~(\ref{eq:spectrum}) were allowed to vary  
when fitting the spectral shape with the time-integrated 
data of the whole period. For the light curve measurements, the spectral 
parameters $\Gamma$ and $E_0$ were kept constant and only the 
normalization $F$ was left free to vary in the likelihood maximization, since 
the counts during a single transit are often not sufficient for a 
multi-parameter fit to converge.

In the light curves shown in the results section we include all 
flux measurements and their uncertainties (1 standard deviation), even if they 
do not constitute a 
significant detection by themselves. 
The likelihood-maximization procedure can produce negative flux 
normalizations. These are obviously non-physical as gamma-ray flux measurements 
but occur when low statistics lead to an underfluctuation of the 
event count compared to the background estimate in a sufficient number 
of analysis bins.

\subsection{Variability Analysis}
\label{sec:variability}

\subsubsection{Likelihood Variability Test}
\label{llhvar}
The maximum-likelihood approach is also used to test if the daily 
flux measurements in a light curve are consistent with a source flux that is 
constant in time over the whole period under consideration. 
We consider the 
likelihood $\mathcal{L}_i(M)$ for the observation in time interval $i$ 
under two different assumptions for the hypothesis $M$:
\begin{itemize}
\item $\mathcal{L}_i(F_i)$, where $F_i$ is the best-fit flux value for time 
interval $i$, as obtained from a likelihood maximization with only this flux as 
a free parameter. The light curves show these flux values $F_i$.
\item $\mathcal{L}_i(F_{\mathrm{const}})$, where $F_{\mathrm{const}}$ is the 
best-fit flux value 
for the time-integrated data set, as obtained from a likelihood 
maximization with only this flux as a free parameter.
\end{itemize}
These definitions allow us to compare the likelihood of individual flux 
measurements with that of a constant flux.
Similar to Section 3.6 of~\citet{2FGL}, we define a test statistic as
twice 
the differences between the logarithms of these likelihood values, summed 
over all intervals:
\begin{equation}
\label{eq:TSvar}
 \mathrm{TS}_{\mathrm{var}} =  2 \sum_i{ \left( \ln{\mathcal{L}_i(F_i)} - 
\ln{\mathcal{L}_i(F_{\mathrm{const}})} \right) } \quad .
\end{equation}
If the null hypothesis of a constant flux is true, then the 
distribution of TS$_{\mathrm{var}}$ 
can be approximated as $\chi^2(n_{\mathrm{dof}}-1)$, according to Wilks' 
theorem~\citep{Wilks1938}.
By applying this variability test to light curves of empty sky 
locations, we found that $\chi^2(n-1)$ indeed matches the distribution 
of TS$_{\mathrm{var}}$ for random fluctuations around zero if we use an 
effective
$n = 1.06 n_{\mathrm{dof}}$, where $n_{\mathrm{dof}}$ is the number of 
degrees of freedom in the light curve.
We calculate the probability for a given source to be 
consistent with the constant flux hypothesis by integrating the $\chi^2(n-1)$ 
distribution above the TS$_{\mathrm{var}}$ value obtained for 
the light curve of that source.

\subsubsection{Bayesian Blocks}
\label{BB}

If a light curve is variable, we can use the Bayesian blocks 
algorithm~\citep{Scargle2012} to find an optimal 
segmentation of the data into regions that are well represented by a constant 
flux, within the statistical uncertainties.
We adopted the so-called {\it point 
measurements} fitness function for the Bayesian blocks algorithm, described in 
Section~3.3 of~\citet{Scargle2012} and applied it to the daily flux data points 
to find the change points at the transition from one flux state to the next. 
The 
algorithm requires the initial choice of a Bayesian prior, called 
$\mathrm{ncp}_{\mathrm{prior}}$,
for the probability of finding a new change of flux states, where 
$\gamma=\exp{(-\mathrm{ncp}_{\mathrm{prior}})}$ is the constant factor 
defining {\it a priori} how much less likely it is to find $k + 1$ 
change points instead of 
$k$ points. In order to choose this prior, we simulated light curves for random 
fluctuations around a constant flux value and required a false 
positive probability of 5\% for finding one change point.
We found this to 
be fulfilled by adopting $\mathrm{ncp}_{\mathrm{prior}} = 6$. 
We checked that 
varying the number 
of light curve 
points between 400 and 500 as well as using different relative uncertainties 
in the simulation to cover the range of observations for our three sources has 
negligible effect on the derived $\mathrm{ncp}_{\mathrm{prior}}$ value. 
The false positive probability accounts for any internal trials of the 
algorithm and results in a relative frequency of 5\% for identifying a 
change point that is not a true flux state change for each light curve 
\citep[see Section 2.7 of][]{Scargle2012}.
The values of the constant flux amplitude within each block, defined by the 
position of the change points, are the 
averages of the corresponding daily measurements, weighted by the inverse 
square of the individual flux uncertainties.

\subsection{Multiwavelength Correlations}

A detailed comparison of simultaneous multiwavelength 
data for features observed in the HAWC light curves is beyond the scope of 
this paper. Instead, we present a first look at multi-instrument 
comparisons of unbiased, 
long-term monitoring that, like the HAWC data, provide 
daily binning and are not affected by seasonal visibility or weather-related 
gaps. Public data with comparable sampling and duty cycle for observations of 
Mrk 421 
and Mrk 501, in particular no gaps larger than a few days, are currently only 
available from very few other monitoring instruments. 
We checked the lower energy gamma-ray light curves with daily binning from the 
Fermi Large Area Telescope (LAT) Monitored Source 
List\footnote{The \textit{Fermi-LAT Monitored Source List Light Curves} can be 
found at 
\url{http://fermi.gsfc.nasa.gov/ssc/data/access/lat/msl_lc}.}, 
with an energy coverage from 100~MeV to 
300~GeV.
For neither Mrk 421 
nor Mrk 501 strong flares were detected on a 1-day timescale and none of the 
daily-averaged integral fluxes exceeded a typical Fermi-LAT alert threshold of 
$10^{-6}$\phr.
These results were generated by an automated analysis pipeline and a re not 
suitable for detailed comparisons of absolute fluxes. A dedicated analysis and 
the study of the correlation between the high energy emission detected by 
Fermi-LAT and the HAWC TeV data will be presented in a forthcoming publication.

In the X-ray band, we can compare our data to the daily light curves 
provided by the Swift/Burst Alert Telescope 
(BAT)~\citep{SwiftBAT2013}. This instrument covers energies between 15 and 50 
keV and, for catalog sources like those discussed here, has a median exposure 
of 1.7 hours per day that can vary throughout the year but stays $<5.4$~hours 
for 95\% of the days.
The Swift-BAT light curves are sampled with one data 
point per day, based on Modified Julian Dates (MJD), and are thus not perfectly 
aligned with the 
binning in local sidereal days that was chosen as a natural frequency for the 
HAWC data. The observations are also not necessarily exactly simultaneous on 
the scale of hours.

\subsection{Systematic Uncertainties}
\label{systematics}

A detailed analysis of systematic uncertainties of gamma-ray fluxes measured 
with HAWC in~\citet{HawcCrab} concludes with estimating a $\pm50$\% 
uncertainty in 
the flux normalization. 
This uncertainty affects all per-transit flux 
measurements only as a common change in absolute scaling and thus does not 
impact the relative magnitude of daily flux measurements. 
The results of 
variability studies and change 
point identification can only be affected by systematic uncertainties that 
change between individual sidereal days or time periods.
The calibration is 
monitored and is very stable. Updates of calibration parameters were only  
performed to accommodate hardware changes, for example additions of PMTs. 
The remaining hardware-related potential source of 
variability is removal or replacement of individual PMTs due to maintenance and 
repairs. This has been found to affect flux measurements by less than $\pm5$\%.
A higher level systematic uncertainty that could in principle affect 
individual fluxes is due to the possibility that the blazar spectra vary with 
time or flux state~\citep[see e.g.][]{Mrk421Krennrich2002}. 
We simulated gamma-ray fluxes with 
different spectral parameters that are allowed within the uncertainties of 
our data and analyzed them with the fixed parameters 
used in the light curve analysis. For each source we determined an optimal 
threshold above which we perform the analytical flux integration by requiring 
that the difference between the photon fluxes for different spectral hypotheses 
is minimal. These threshold values are used when quoting the photon 
fluxes in the results sections: 1~TeV for the Crab Nebula, 2~TeV for Mrk 421, 
and 3~TeV for Mrk 501.  The resulting uncertainty on individual flux 
values under spectral hardening or softening is $\pm5$\%. 
The combination of these two potentially time-dependent systematic 
uncertainties is significantly smaller than the statistical uncertainty of 
the per-transit flux values and thus marginal with respect to the 
analysis of variability features.

We have performed further tests of the 
robustness of the likelihood variability estimation. Using the Crab Nebula 
as a reference, we found that changes in the analysis 
procedure with respect to background estimation (different smoothing 
procedures within the direct integration), data selection (excluding the three 
highest bins with an average $\leq1$ photon per day for our sources), and 
spectral model (power law and log parabola) only changed the resulting 
TS$_{\mathrm{var}}$ 
value of the variability by 
$\leq 0.2$ standard deviations of the null hypothesis. We therefore find no 
indication that the map-making and likelihood analysis can introduce 
significant variability features.
We conclude that our analysis of flux variations and identification of flaring 
states in this paper is not limited by these systematic uncertainties. 

\begin{figure*}[t]
\centering
\includegraphics[width=0.98\textwidth]{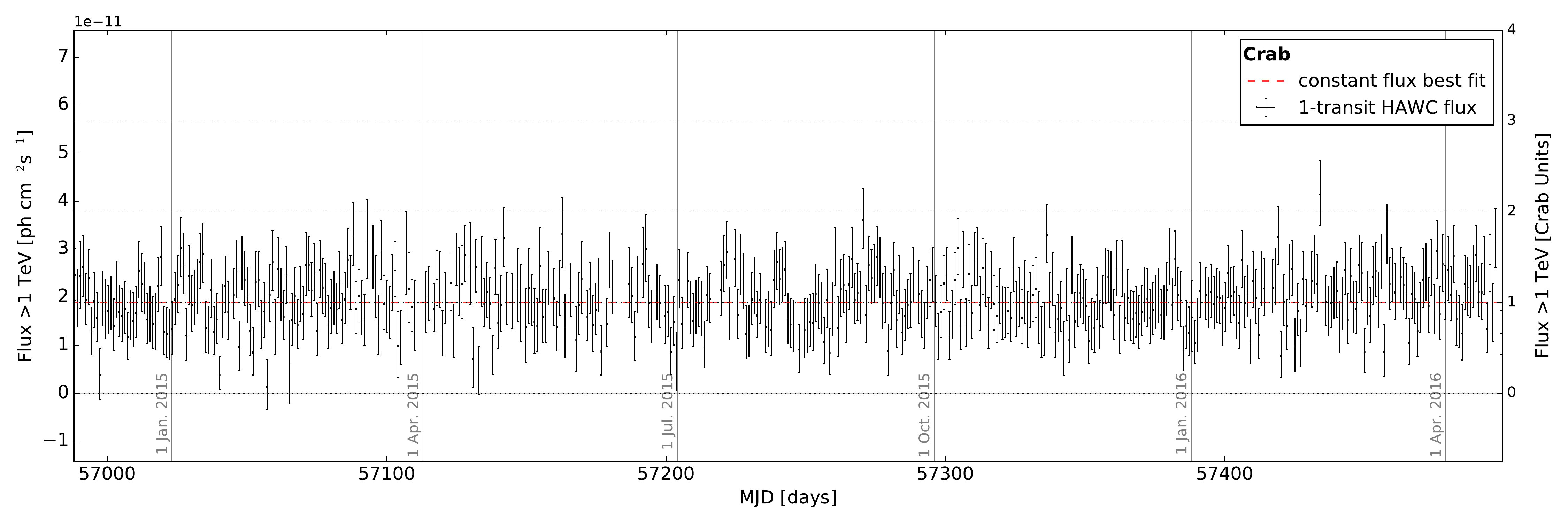}
\caption{Flux light curve for the Crab Nebula, daily sampling for 
472 transits between 2014 November 26 and 2016 April 20. 
The integrated fluxes are derived from fitting $F_i$ in spectral function 
$dN/dE=F_i \left ( E / (1\, \mathrm{~TeV}) \right )^{-2.63} $, with Crab 
Units normalized to the average HAWC flux measured over the whole time 
period. The dashed red line is the flux average when assuming a constant flux 
for the whole period. 
}
\label{fig:lcCrab}
\end{figure*}

For the interpretation of absolute flux values
it is helpful to compare our measurements to a gamma-ray 
reference flux. We therefore convert fluxes to multiples of 
the HAWC-measured Crab Nebula flux \citep[$1.89\cdot 
10^{-11}$~ph~cm~$^{-2}$~s$^{-1}$, see detailed analysis in][]{HawcCrab}
as Crab Units (CU) with a common threshold of 1~TeV. This threshold was 
chosen to provide easier comparisons between the different sources and the 
literature. 
We still have to consider the flux uncertainty introduced by the choice of 
a fixed spectral assumption for which the analytical integration above 1~TeV is 
performed.
We used the time-integrated HAWC data to fit the flux normalization of each 
of the three sources discussed here with a number of different power law 
indices and cut-off values that cover the individual statistical and systematic 
uncertainty range. 
We find that the maximum uncertainty of the photon fluxes in CU due to the 
spectral assumption is $\pm25$\%.

\section{Results for the Crab Nebula}
\label{sec:crab}

\subsection{Flux Light Curve}

The Crab Nebula is the brightest galactic TeV point source. A detailed 
analysis of 
time-integrated HAWC data for this source is presented in~\citet{HawcCrab}. In 
Fig.~\ref{fig:lcCrab} 
we show the results of applying the likelihood analysis to the sidereal day 
maps at the location of 
the Crab Nebula.
We use a fixed spectrum with index $\Gamma=2.63$ in 
equation~(\ref{eq:spectrum}) and no 
exponential cut-off, $E_0\rightarrow\infty$, based on the best 
fit value 
obtained in the HAWC catalog~\citep{HawcCatalog2016}. 

The left-hand y-axis in Fig.~\ref{fig:lcCrab} indicates the photon flux (\phr) 
after analytically 
integrating 
the spectrum above 1~TeV for the best fit normalization. The right-hand axis 
shows the Crab Units (CU)
defined by dividing the flux by the time-averaged HAWC measurement of the Crab 
flux, 
also indicated as a dashed line in the figure.
We use Modified 
Julian Dates (MJD) for labeling the time axes and highlight the duration of 
HAWC measurements (6 sidereal hours) through horizontal bars.

We applied the variability test outlined in Section~\ref{sec:variability} to 
the light curve with 
1-transit intervals and found a $\mathrm{TS}_{\mathrm{var}} = 517.9$, 
with a probability of 0.292 ($1.1$ 
standard deviations) of measuring the same or a larger TS value for a 
constant flux hypothesis. An 
analysis of the light curve with 
the Bayesian blocks algorithm with a false positive probability of 5\% 
reveals no change points. HAWC daily flux measurements thus show no 
indication of variability in data from the Crab Nebula.

\begin{figure}[t]
\centering
\includegraphics[width=0.47\textwidth]{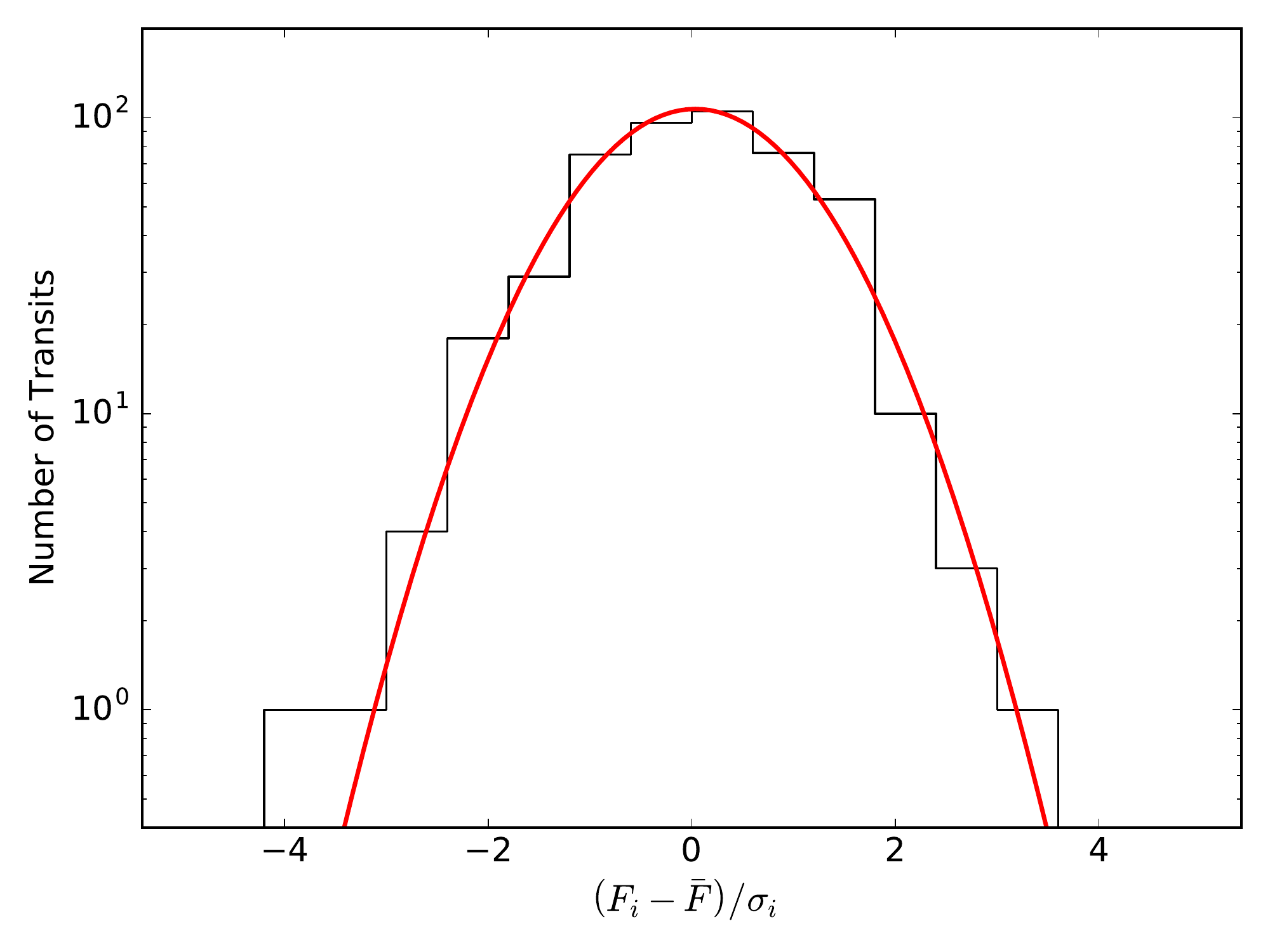}
\caption{Histogram of the differences between per-transit fluxes $F_i$ from 
the light curve (Fig.~\ref{fig:lcCrab})
and the average Crab Nebula flux $\bar{F}$, divided by the uncertainties 
$\sigma_i$. 
The distribution is well described by a fit with a Gaussian function and the 
fitted parameters are consistent within uncertainties with a center at zero and 
a width of one. }
\label{fig:histFluxCrab}
\end{figure}

In Fig.~\ref{fig:histFluxCrab} we show a histogram of 
$(F_i-\bar{F})/\sigma_i$, where $F_i$ and $\sigma_i$ are the fluxes and 
uncertainties from Fig.~\ref{fig:lcCrab} and  $\bar{F}$ is the best fit value 
for a constant flux. A fit to a Gaussian function yields a center at $0.035 \pm 
0.050$ and a width of $1.033 \pm 0.036$, confirming that the observed flux 
distribution is consistent with arising from a constant source flux.

\subsection{Discussion}
Based on measurements by other instruments, the 
Crab is generally believed to 
be a steady source\footnote{We 
ignore here the very high energy pulsed emission directly 
from the pulsar that is very weak compared to the pulsar wind nebula's 
emission.} at TeV 
energies~\citep{VeritasCrabFlare, HessCrabFlare, 
ArgoCrab}.
The non-detection of variability in TeV emission from the Crab Nebula with 
HAWC is in agreement with these results. 
We conclude that HAWC daily light curve measurements and the likelihood 
variability check are a robust test of the steady gamma-ray 
source hypothesis and that any systematic uncertainties in the HAWC data 
are very unlikely to mimic significant variability in this analysis.

Given that the Crab Nebula is known to flare in lower energy 
bands, we can use the unique daily TeV light curve data to 
constrain any TeV flux enhancement during such episodes. During the 17 months 
included here, the Fermi-LAT collaboration reported an increased 
gamma-ray flux for energies $>100$~MeV between 2015 
December 28 and 2016 January 9, reaching up to 
$\sim 1.7$ times the average flux~\citep{ATEL8519}. 
The maximum HAWC 1-transit flux during this period was $ (2.14 \pm 0.54)\cdot 
10^{-11}$\phr above 1 TeV on 2016 January 7, only 0.46 standard deviations 
above the average flux. This corresponds to an upper limit at 95\% confidence 
level of $ 3.04 \cdot 10^{-11}$\phr above 1 TeV, 1.6 times the average flux.
When we combine the HAWC measurements over 
the 12 transits\footnote{Two out of 14 transits did not pass the 
quality criterion of $>50$\% transits coverage.} included in this period, 
we obtain a flux 
measurement of $ (1.42 \pm 0.15)\cdot 10^{-11} $\phr above 1 TeV, 0.75 times 
the average flux and consistent with a random fluctuation. 
We conclude that we observe no significant change in the TeV flux during this 
MeV flare period.

\section{Results for Mrk 421}
\label{sec:mrk421}

\subsection{Source Characteristics}
\label{sec:mrk421char}

Mrk 421 is a BL Lacertae type blazar with a redshift of 
$z = 0.031$~\citep{2MASSz}. It was the first extragalactic object 
discovered in the TeV band~\citep{Punch1992} and has been extensively 
studied by many TeV gamma-ray observatories. Mrk 421 
is known to exhibit a high degree of variability in its emission and yearly 
average fluxes are known to vary between 
a few tenths and $\sim1.9$ times the flux of the Crab
Nebula~\citep{Veritas14yearsMrk421}. 
Variability has been observed down to time scales of hours or less and its 
spectral 
shape is known to vary with its brightness~\citep{Mrk421Krennrich2002}.

By using the time-integrated HAWC data for Markarian 421, we fit the spectral 
shape with the 
likelihood methods discussed in Section~\ref{sec:FluxSpectral}, leaving the 
normalization $F$, the photon index $\Gamma$, and 
the exponential cut-off $E_0$ free. The resulting best fit values are
$F =  (2.82 \pm 0.19_{\mathrm{stat}} \pm 1.41_{\mathrm{sys}} ) \cdot 
10^{-11}$\unorm for the time-averaged normalization at 1~TeV, a photon index
$\Gamma=2.21 \pm 0.14_{\mathrm{stat}} \pm 0.20_{\mathrm{sys}}$ and an 
exponential cut-off at $E_0=5.4 \pm 1.1_{\mathrm{stat}} \pm 
1.0_{\mathrm{sys}}$~TeV. The significance of this description over the 
background-only hypothesis is $\mathrm{TS} = 1232.47$ or 35.1 standard 
deviations. When compared to a pure power law hypothesis ($E_0 
\rightarrow \infty$), the fit with a cut-off is clearly the better 
description, preferred at $\Delta \mathrm{TS}=64.8$ or 8 standard deviations.
We use these values for index and cut-off as a fixed set of 
parameters when fitting the flux normalization for each sidereal day, reported 
in the following section.

\subsection{Flux Light Curve}
\label{sec:mrk421lc}

\begin{figure*}[t]
\centering
\includegraphics[width=0.98\textwidth]{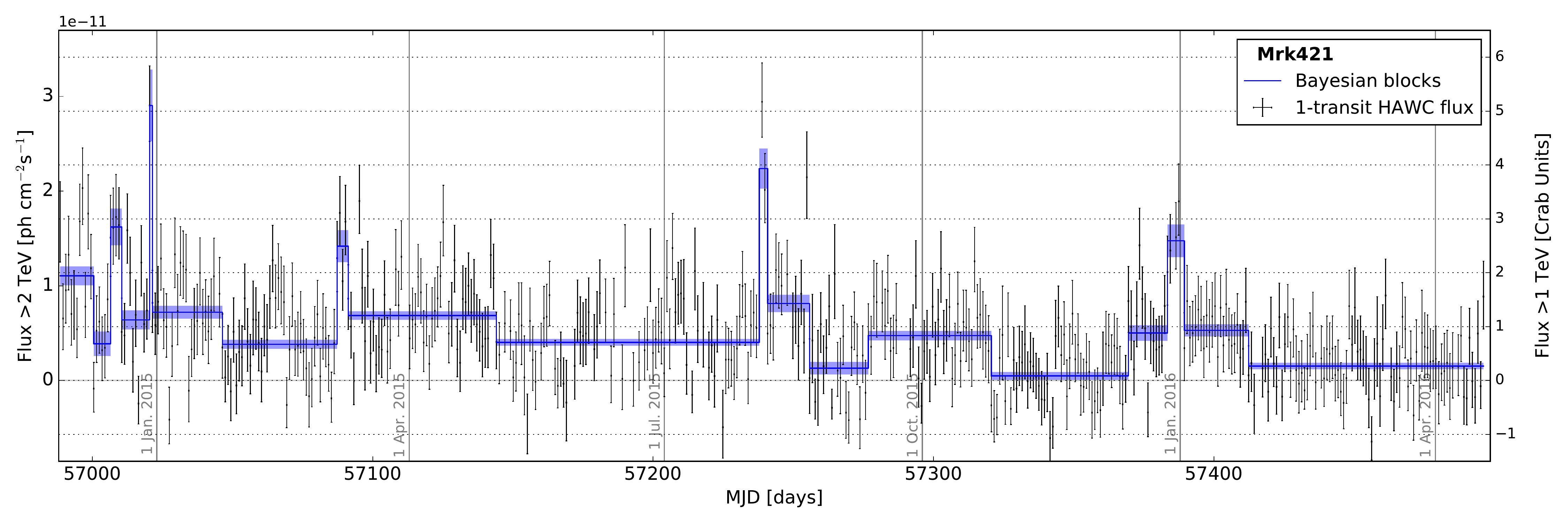}
\caption{Flux light curve for Mrk 421 with sidereal-day sampling for 471 
transits 
between 2014 November 26 and 2016 April 20. The integrated fluxes are
derived from fitting $F_i$ in 
$dN/dE=F_i \left ( E/(1\, \mathrm{TeV})\right )^{-2.2}  
\exp{\left(-E/(5\, \mathrm{~TeV})\right)} $ 
and converted to Crab Units via dividing by the HAWC measurement of the 
average Crab Nebula gamma-ray flux. 
The blue lines show the distinct flux states between change points identified 
via the Bayesian blocks analysis with a 5\% false positive probability.
}
\label{fig:lcMrk421}
\end{figure*}

The flux light curve for Mrk 421 with 1-transit intervals is shown in 
Fig.~\ref{fig:lcMrk421} .
Photon flux units (left y-axis) are based on analytical 
integration of the fixed spectral shape above a threshold of 2~TeV that 
minimizes flux uncertainties due to spectral variations. 
The conversion to 
CU (right axis) is based on average HAWC Crab 
Nebula measurements, see Section~\ref{sec:crab}, for a common threshold of 
1~TeV in order to allow comparisons between the different sources.
The average flux for the 17 months period is determined via a fit of the 
combined data under a constant flux assumption and yields $(4.53 \pm 
0.14)\cdot 10^{-12}$\phr above 2 TeV.

Applying the likelihood variability test to this light curve yields 
TS$_{\mathrm{var}} = 1154.9$, 
which corresponds to a p-value $4.40\cdot 10^{-54}$ based on the expected 
$\chi^2$ distribution for constant flux models and clearly shows the variable 
nature of the TeV emission from Mrk 421.
The highest per-transit flux value, $(2.94 \pm 0.37) \cdot 10^{-11}$ 
ph~cm~$^{-2}$~s$^{-1}$ above 2~TeV was measured for 
 MJD 57238.74 -- 57238.99 (2015-08-04 UTC 17:40 -- 23:40), with a pre-trial 
significance of 9.3 standard deviations compared to the null hypothesis.
A flux that is only slightly lower, $(2.91 \pm 0.38) \cdot 10^{-11}$ 
ph~cm~$^{-2}$~s$^{-1}$, was observed during MJD 57020.33 -- 57020.58 
(2014-12-29 UTC 8:00 -- 14:00) and highlights that the variability occurs 
on time scales of less than one day, since the flux value for the day before 
and after this maximum are a factor of $\sim3$ and $\sim4$ lower, 
respectively.

The Bayesian blocks algorithm with a prior corresponding to a false 
positive probability of 5\% identifies 18 change points in the light curve 
shown in 
Fig.~\ref{fig:lcMrk421}. The flux amplitudes for the periods between two 
change points that are consistent with a constant flux are included as 
blue lines with a shaded region for the statistical uncertainty of one standard 
deviation. These block positions and amplitudes are listed in 
Table~\ref{tab:bbTableMrk421} in the Appendix.

\subsection{Discussion}

The spectral fit results, $\Gamma=2.21 \pm 0.14_{\mathrm{stat}} \pm 
0.20_{\mathrm{sys}}$ and $E_0=5.4 \pm 
1.1_{\mathrm{stat}} \pm 1.0_{\mathrm{sys}}$~TeV, are consistent with 
spectral shapes previously observed~\citep[see e.g.][]{Albert2007Mrk421}.
If we compare to the range of VERITAS spectral fits as a function of flux state 
in~\citet{Veritas2011Mrk421}, we find that the average HAWC spectrum is closest 
to the parameters for the \textit{Mid-state} level, $\Gamma^{\mathrm{M}}=2.278 
\pm 0.037$ and $E_0^{\mathrm{M}} = 4.36 \pm 0.58$~TeV.
A more detailed analysis of the HAWC spectral fits and a discussion of the 
absorption features of the EBL is beyond the scope of this 
paper. We will revisit this in a separate paper and take advantage of better 
energy estimation techniques for HAWC data that are currently under development 
to enhance the sensitivity to the curvature at the highest energies.

\begin{figure}[t]
\centering
\includegraphics[width=0.47\textwidth]{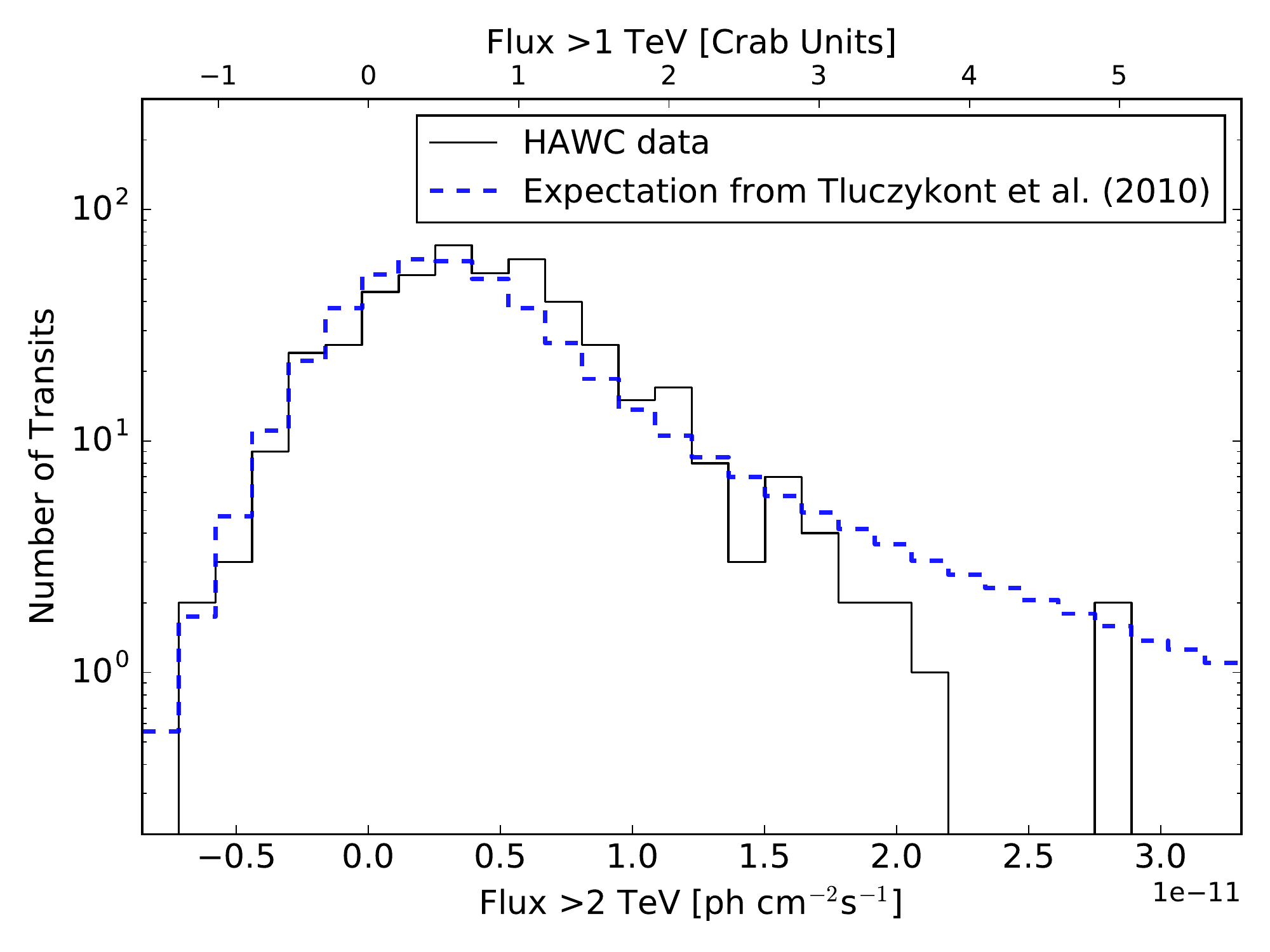}
\caption{Histogram of the 1-transit flux measurements for Mrk 421. It is 
compared to an averaged histogram (blue, dashed) 
of fluxes drawn from a function that fits the distribution of archival Mrk 
421 flux states from ~\citet{Tluczykont2010}, see text for details.}
\label{fig:histFluxMrk421}
\end{figure}

HAWC data can confirm and track the variability of Mrk 421 via daily flux 
measurements.
By applying the Bayesian blocks algorithm, we identified 19 distinct 
flux states. The apparent substructure within 
some blocks in Figure~\ref{fig:lcMrk421} is likely to be
due to flux variations on shorter time scales which 
cannot be resolved by the present analysis, given the predetermined 5\% false 
positive probability and the uncertainties of our measurements. As a stability 
check, we lowered the false positive condition from 5\% to 10\%, which led 
to 
the identification of only one additional block around MJD 57065. 

In Fig.~\ref{fig:histFluxMrk421}, a histogram of all flux measurements 
from Mrk 421 highlights the spread of the observed flux states. 
We can compare this distribution to a function from
\citet{Tluczykont2010} that was derived as a good fit to archival data, 
composed of the sum of a normal distribution ($f_{\mathrm{Gauss}}$) around a 
low flux peak and a log-normal part ($f_{\mathrm{LogN}}$) describing a tail to 
higher fluxes. 
The number of observations as a function of the flux 
$x$ is given by:
\begin{eqnarray}
 \label{eq:Tlucz}
 f_{T}(x) &=& f_{\mathrm{Gauss}} + f_{\mathrm{LogN}}  \\
          &=& \frac{n_{\mathrm{Gauss}}}{\sigma_{ \mathrm { Gauss}}\sqrt{2\pi}} 
\exp{\left(-\frac{(x-\mu_{\mathrm{Gauss}})^2}{2\sigma_{ \mathrm { Gauss}}^2 } 
\right) } \nonumber \\
 && +
\frac{n_{\mathrm{LogN}}}{x \sigma_{ \mathrm { LogN}}\sqrt{2\pi}} 
\exp{\left(-\frac{(\ln(x)-\mu_{\mathrm{LogN}})^2}{2\sigma_{\mathrm{LogN}}^2
} \right) }
\;, \nonumber
\end{eqnarray}
with the best-fit parameters 
$n_{\mathrm{Gauss}} = 48.08$ ,
$\mu_{\mathrm{Gauss}} = 0.3285$ , 
$\sigma_{\mathrm{Gauss}} = 0.1137$ ,
$n_{\mathrm{LogN}} = 45.55$ ,  
$\mu_{\mathrm{LogN}} = 0.1025 $ ,
and $\sigma_{\mathrm{LogN}} = 1.022 $ . 
Here we follow the convention from the reference of measuring $x$ in CU above 
1 TeV but setting its unit to 1 in the formula.
In order to account for the HAWC measurement uncertainties, we use a two-step 
process to define samples that each have 471 flux 
values, matching the size of the data set. First, we draw 471 random values 
according to the distribution in equation~(\ref{eq:Tlucz}) and use these value 
as centers and 
the standard deviation values from data as widths to define 
471 normal distributions. Then, we draw one random value form each of these 
normal distributions to obtain a set of fluxes that reflects the uncertainties 
of the HAWC data. We average over 10,000 such samples to obtain 
the prediction for a HAWC measurement of this flux distribution. It is included 
in Fig.~\ref{fig:histFluxMrk421} (blue, dashed line).
We compare this expectation with the 
histogram of HAWC data via a Kolmogorov-Smirnov (KS) test and find a 
probability of 
$0.0008$ that they arise from the same distribution. This value is 
stable under changes of the histogram binning and rescaling the HAWC 
fluxes within the systematic uncertainty of $\pm25$\% leads to a maximum KS 
probability value of $0.0046$. Since we account for HAWC flux 
uncertainties in 
our sampling procedure, we also tested reducing
$\sigma_{\mathrm{Gauss}}$ from~\citet{Tluczykont2010} under the 
assumption that it is mostly reflecting  
measurement uncertainties in the fitted data, but found only 
smaller KS probabilities. 
Since equation~(\ref{eq:Tlucz}) is based on the fit to 
a compilation of measurements from many different 
instruments, it is hard to assess the
systematic uncertainties of this parametrization. We have to consider that 
the large gaps in time coverage and a likely bias due to observations 
triggered by multiwavelength alerts for the public data 
in~\citet{Tluczykont2010} can lead to a 
fit that does not well represent the average daily flux distribution 
for Mrk 421. We compare this function here for the first time with data from an 
unbiased, regular monitoring with a single, stable detector and conclude that 
these 17 
months of HAWC observations cannot be well described by 
equation~(\ref{eq:Tlucz}). 
The current level of statistical uncertainties of the HAWC data prevents us 
from obtaining a stable fit of the 6 parameters from 
equation~(\ref{eq:Tlucz}) or testing if the tail 
of higher fluxes is indeed best described with a log-normal distribution which 
could indicate an origin of variability from multiplicative processes. 
Increased statistics and the 
coverage of more high flux states with new HAWC data will provide the basis for 
obtaining a better functional description of Mrk 421 flux states.

Since February 2016, the daily flux measurements for both 
Mrk 421 and Mrk 501 have been automatically performed at the HAWC 
site immediately after the end of each transit. The preliminary analysis is 
based on the so-called \textit{online} reconstruction, performed with only a 
few seconds' time lag on all recorded events and a preliminary calibration and 
data quality selection. Our threshold for issuing alerts about flaring 
states for both sources is a flux value equivalent to 3~CU in a single transit, 
which corresponds to a detection at $\sim5$ standard deviations for Mrk 421. 
In the 17 
months of data included here, Mrk 421 surpassed this threshold during 11 
transits, 2.3\% of the time.

\subsection{Multiwavelength Comparisons}
\label{sec:mrk421mw}

\begin{figure*}[t]
\centering
\includegraphics[width=0.98\textwidth]{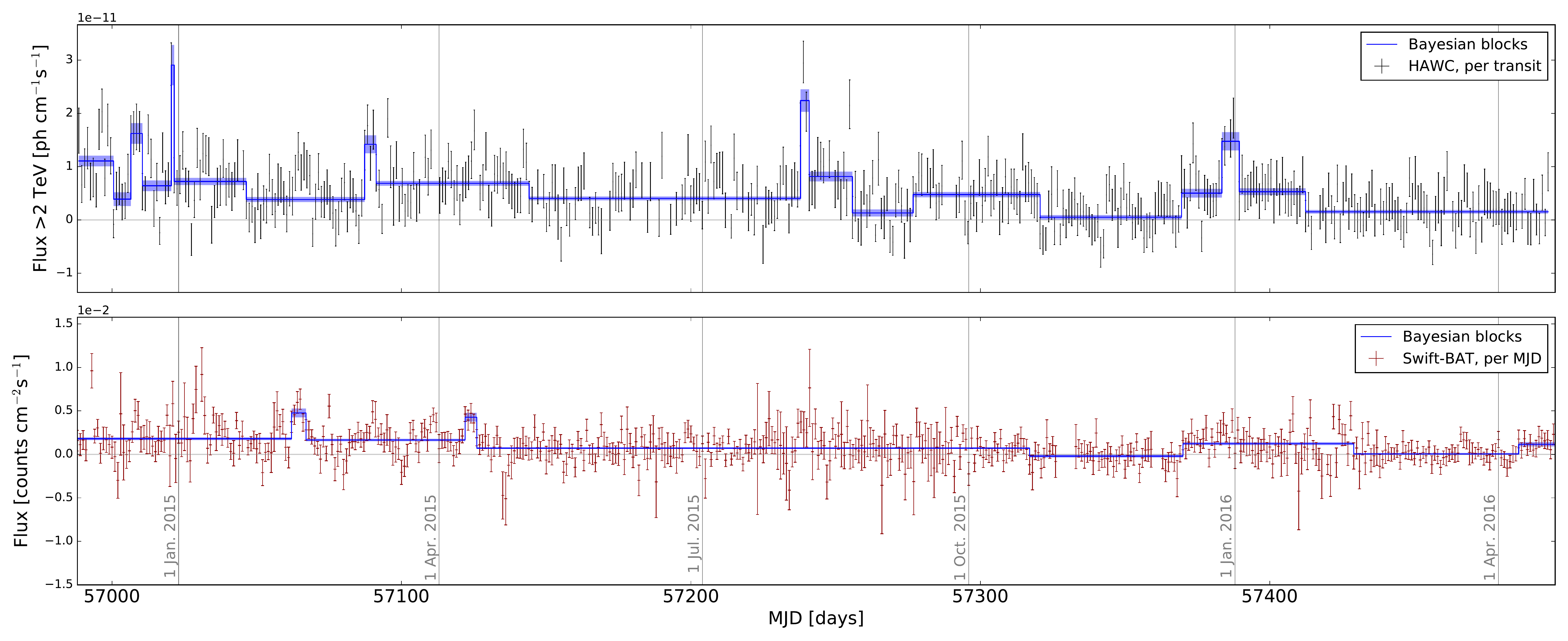}
\caption{Comparison of light curves for Mrk 421 between data from HAWC ($>2$ 
TeV, top panel) and Swift-BAT 
(15 to 50 keV, bottom panel). The results 
of the Bayesian blocks analysis with a 5\% false positive probability are 
included (blue lines).}
\label{fig:Mrk421MW}
\end{figure*}

In Fig.~\ref{fig:Mrk421MW}, we compare the HAWC TeV measurements to light 
curves from the Swift-BAT (15 -- 50~keV)\footnote{Public light curve data from 
\url{ http://swift.gsfc.nasa.gov/results/transients/}.}
monitoring instrument that provides very similar sampling and instrument duty 
cycle. 
The Swift-BAT data allow us to apply the same 
Bayesian blocks algorithm as used for HAWC data. The ratio of average 
error to mean flux value is larger for Swift-BAT light curves than for the 
HAWC data, but simulations show that the same Bayesian prior 
value ($\mathrm{ncp}_{\mathrm{prior}}=6$) will guarantee the same false 
positive 
probability of 
5\%. This analysis identifies 8 change points, i.e. 9 blocks, in the X-ray data.
The only Swift-BAT flux state that matches one of the HAWC flux states with 
less than 10 days' difference in start and end times is the lowest one
\footnote{The Swift-BAT weighted mean 
amplitude for this block is negative but compatible with zero within 1.1 
standard deviations.}. 
None of the highest HAWC-measured flaring states are 
mirrored in the blocks for the X-ray light curve.
We cannot exclude that the size of the statistical 
uncertainties hides correlated features at the day 
scale, considering that at least the Swift-BAT energy band seems to cover 
mostly a steeply falling part of the spectral energy distribution observed in 
the past~\citep{FermiMrk421SED}. 
It is possible that this is 
due to insufficient overlap between the instruments' exposures during one 
day, since we established significant TeV variability 
within less than one sidereal day in Section~\ref{sec:mrk421lc}. Similar cases 
of missing correlations for bright TeV flares have been observed 
before~\citep[see e.g.][]{Veritas2011Mrk421,Blazejowski2005}.
On the other hand, when we compare all daily averaged fluxes by calculating the 
Spearman rank correlation coefficient for the HAWC and the Swift-BAT data we 
find a positive correlation of $0.341 \pm 0.030$. The probability for this 
result to occur for uncorrelated data sets of the same size is $3\cdot 
10^{-13}$ and we can thus qualitatively confirm previous 
observations of TeV-to-keV correlations based on (partially biased) IACT 
data~\citep{Fossati2008,Blazejowski2005,Albert2007Mrk421,Horan2009,
Tluczykont2010} .

The only public notification about flaring activity for Mrk 
421 that was sent during the period under investigation is an Astronomer's 
Telegram~\citep{ATEL7654} about an increased X-ray flux observed with 
Swift-XRT (0.3 -- 10 keV) between June 8 and 
June 16, 2015. 
Our Bayesian blocks analysis does not identify any significant flux state 
changes within 1.2 months around these dates.

\section{Results for Mrk 501}
\label{sec:mrk501}

\subsection{Source Characteristics}

Mrk 501 is a BL Lacertae type blazar that is similar to Mrk 421, given its 
distance of $z=0.033$~\citep{2MASSz} and its 
classification as a high-peaked BL Lac object. It is the second extragalactic 
object that was discovered at TeV energies~\citep{Weekes1996}. 
Various studies at TeV energies have shown different features
of low flux states emission and extreme outbursts, for example 
in~\citet{quietMrk5012011}.

Our initial fit of the spectral shape uses the integrated 17 months of 
HAWC data.
When we do not allow curvature in the spectral model, $E_0\rightarrow \infty$ 
in equation~\ref{eq:spectrum}, we obtain the 
best fit values $F =  (4.50 \pm 0.28_{\mathrm{stat}} \pm 
2.25_{\mathrm{sys}} ) \cdot 10^{-12}$\unorm for normalization at 1~TeV and a 
photon 
index $\Gamma=2.84 \pm 0.04_{stat} \pm 0.20_{sys}$. This is consistent 
with results reported in~\citet{quietMrk5012011} and \citet{FermiEraMrk501}.
Leaving also the exponential cut-off free yields a normalization
$F =  (4.40 \pm 0.60_{\mathrm{stat}} \pm 2.20_{\mathrm{sys}} ) \cdot 
10^{-12}$\unorm, 
a photon index $\Gamma=1.60 \pm 0.30_{\mathrm{stat}} \pm 0.20_{\mathrm{sys}}$, 
and an exponential cut-off value of $E_0=5.7 \pm 1.6_{\mathrm{stat}} \pm 
1.0_{\mathrm{sys}}$~TeV. The 
latter result is clearly preferred by $\Delta \mathrm{TS} = 48.64$ or 7.0 
standard 
deviations over the power law fit without a cut-off. Its significance compared 
to the background-only hypothesis is $\mathrm{TS} = 610.49$ or 24.7 standard 
deviations.
For the flux normalization fits performed to construct the flux light curve we 
use the description with the cut-off and keep the 
index and cut-off parameters fixed at the HAWC-measured values.

\subsection{Flux Light Curve}

\begin{figure*}[t]
\centering
\includegraphics[width=0.98\textwidth]{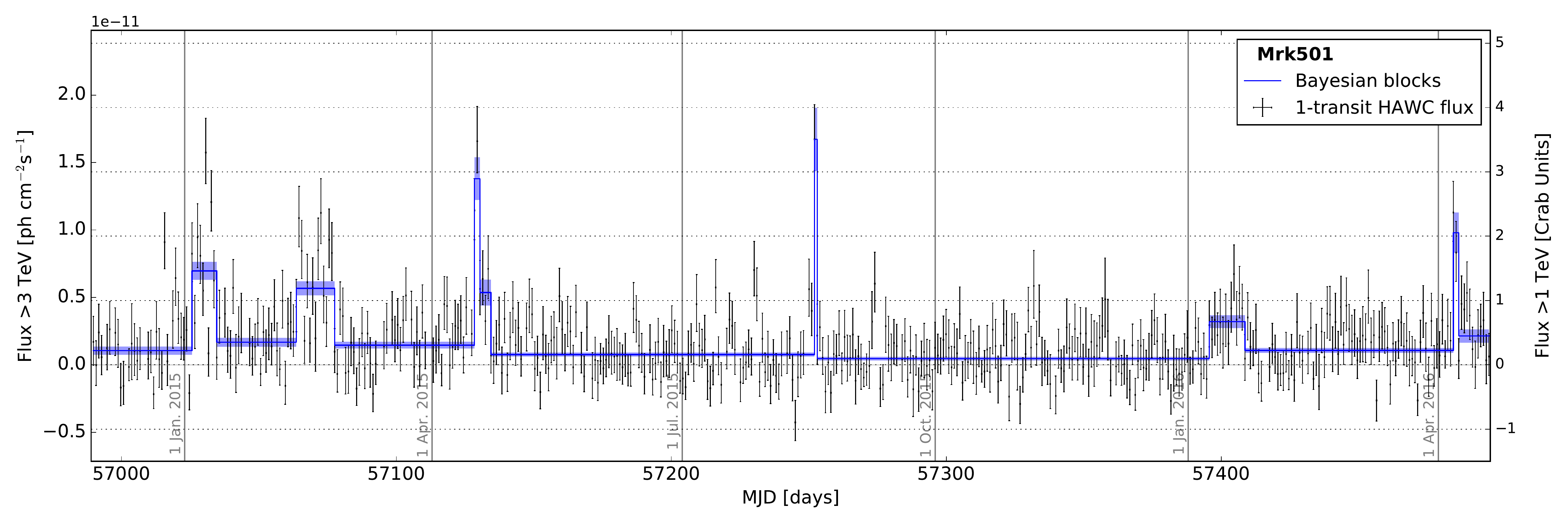}
\caption{Flux light curve for Mrk 501 with sidereal-day sampling for 479 
transits 
between 2014 November 26 and 2016 April 20. The integrated fluxes 
are derived from fitting $F_i$ in  
$dN/dE=F_i \left ( E/(1\,\mathrm{TeV})\right )^{-1.6}  
\exp{\left(-E/(6\,\mathrm{TeV})\right)} $ and converted to Crab Units by 
dividing 
by the HAWC measurement of the average 
Crab Nebula gamma-ray flux. 
The blue lines show the distinct flux states between change points identified 
via the Bayesian blocks analysis with a 5\% false positive probability.
}
\label{fig:lcMrk501}
\end{figure*}

The Mrk 501 flux light curve for all 1-transit intervals that have
$>50$\% coverage with HAWC is shown in 
Fig.~\ref{fig:lcMrk501}. The photon flux is calculated as the analytical 
integration above 3~TeV, the optimal threshold value for Mrk~501 in order
to minimize the systematic uncertainties of the flux measurement due to 
the fixed spectral assumption.
For the 17 months period included here, we find an average flux of $(1.74 
\pm 0.08) \cdot 10^{-12}$\phr above 3~TeV.

The result of the likelihood variability calculation for Mrk 501 is 
TS$_{\mathrm{var}} = 1115.4$, corresponding to a p-value $9.18\cdot10^{-48}$, 
and thus 
clearly establishes variability of the TeV emission measured here.
The highest daily flux, $(1.67 \pm 0.23) \cdot 10^{-11}$\phr above 3 
TeV, was observed during the transit MJD 57251.94 -- 57252.19 
(2015-08-17 UTC 22:40 to 2015-08-18 UTC 4:40) with a pre-trial significance of 
9.5 standard deviations compared to the null hypothesis. 
This is approximately a factor 10 higher than the constant flux fit average and 
shows a 
variability time scale of less than 
one day, since the flux is higher by a factor $\sim4$ compared to the previous 
transit and by a factor $\sim8$ compared to the next transit.

In order to find significant flux state changes in this light 
curve, we applied the Bayesian blocks algorithm.
Given the prior for 5\% false positive probability, the algorithm 
identifies 
13 change points. The amplitudes of periods between these change points are 
consistent with a constant flux and are shown as blue lines in 
Fig.~\ref{fig:lcMrk501}, with shaded 
bands indicating one standard deviation around the mean 
amplitude.
These block positions and amplitudes are listed in 
Table~\ref{tab:bbTableMrk501} in the Appendix.

\begin{figure}[t]
\centering
\includegraphics[width=0.47\textwidth]{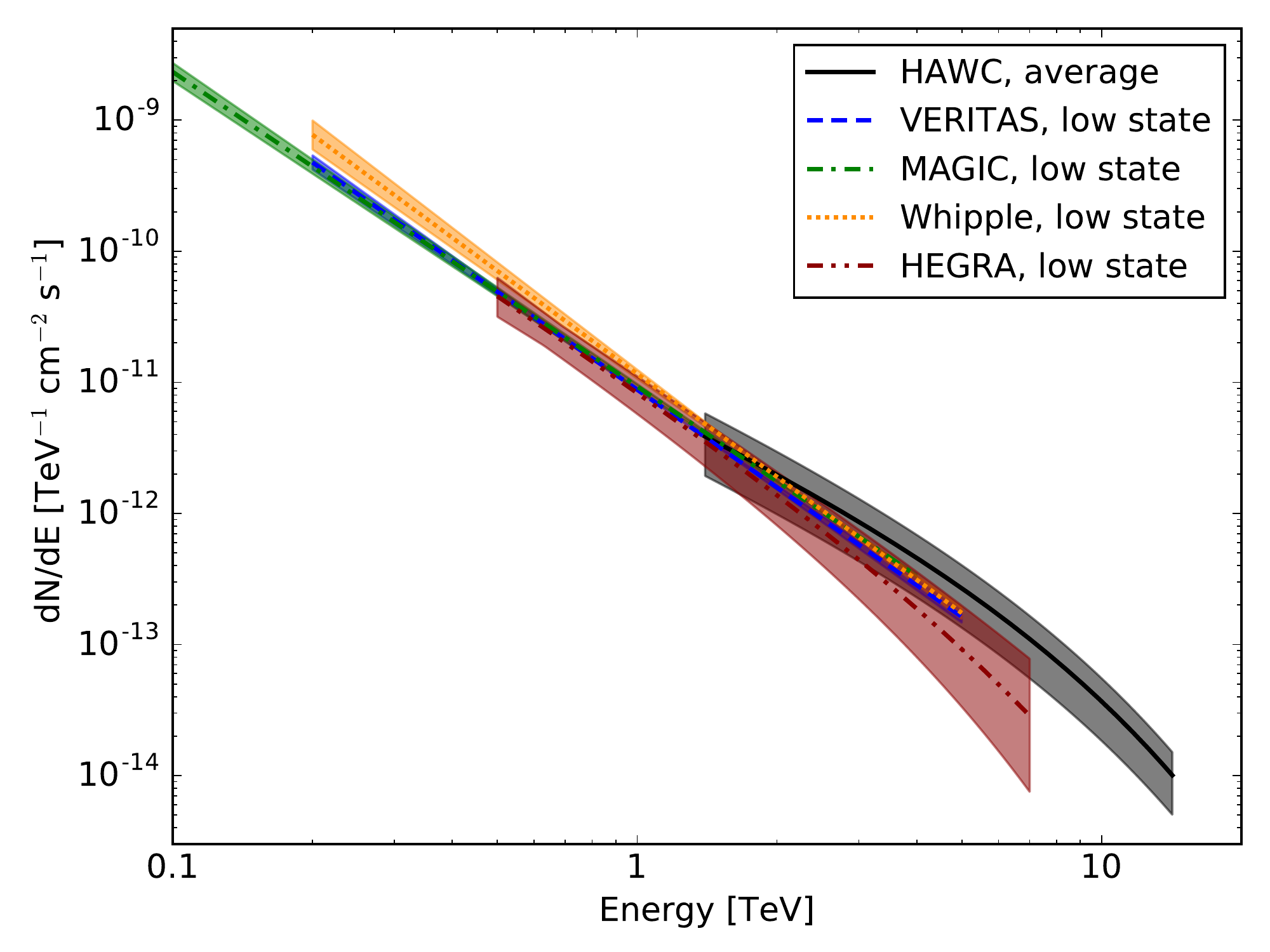}
\caption{The HAWC spectral fit result for Mrk 501 (black) with a band showing 
the statistical and systematic uncertainty range. We compare this fit to 
spectral measurements during low flux states of Mrk 501 with VERITAS 
(blue) and Whipple (orange) from~\citet{Aliu2016}, 
MAGIC (green) from~\citet{Ahnen2017}, and 
HEGRA (brown) from~\citet{Aharonian2001}. These four IACT spectral fits are 
shown in the energy ranges given in the references with bands 
covering statistical uncertainties.}
\label{fig:spectrumMrk501}
\end{figure}

\subsection{Discussion}

The integrated HAWC data for Mrk 501 is best described via a curved spectrum 
that we model with a photon index $\Gamma=1.6$ and an exponential cut-off at  
$E_0=5.7$~TeV. In Fig.~\ref{fig:spectrumMrk501} we compare this result with 
spectra measured by MAGIC~\citep{Ahnen2017}, VERITAS and 
Whipple~\citep{Aliu2016}, as well as HEGRA~\citep{Aharonian2001}. 
We include only the values designated as \textit{low-state} by these 
observatories 
since the HAWC light curve shows long periods of low activity that dominate 
our averaged measurement. The analyses of spectra during flaring states in the 
same publications show generally harder photon indices. The HAWC 
spectrum averaged over 17 months is consistent with these measurements within 
the statistical and systematic uncertainties as shown in 
Fig.~\ref{fig:spectrumMrk501}. The spectral curvature in our measurement 
manifests itself primarily outside the energy range covered by the IACT 
measurements. 
The HAWC energy range was determined from simulation as the central 
interval containing 90\% of expected signal events for the best-fit 
spectrum. While the strong curvature in the spectrum of Mrk 501 
prevents us from constraining the spectral shape above an energy of $\sim$15~TeV 
with the current analysis, HAWC is generally sensitive to gamma ray energies up 
to $\sim$100~TeV, as discussed in~\citet{HawcCatalog2016}.

The HEGRA analysis also obtains 
a better fit 
with an exponential cut-off than a pure power law. The HEGRA cut-off 
value, $5.1(_{-2.3}^{+7.8})_{\mathrm{stat}}$~TeV, is consistent with the HAWC 
value. 
A cut-off in the energy spectrum can arise from gamma-ray absorption through the 
EBL~\citep[see e.g.][]{Dominguez2011} or originate in processes intrinsic to the 
source, for example, a limit to the energies of injected particles, changes in 
the Klein-Nishina 
scattering cross section~\citep{Hillas1999}, or absorption through photon 
fields 
in the lower jet~\citep{Dermer1994}. 
The best-fit  
photon index $\Gamma=1.6 \pm 0.30_{\mathrm{stat}} \pm 0.20_{\mathrm{stat}}$ 
that we measure is hard compared to, for example, Mrk 421 but still greater 
than the lower limit of $\sim 1.5$ for Fermi-acceleration in 
shocks~\citep{MalkovDrury2001}. 
At lower 
energies, up to $\sim~300$~GeV, spectral hardening with 
photon indices down to $\sim 1.0$ has been observed for Mrk 501 by 
Fermi-LAT~\citep[see e.g.][]{Shukla2016}.
We will provide a more detailed study of the spectral energy distribution and 
EBL absorption for Mrk 501 with HAWC data in a separate publication.

\begin{figure}[t]
\centering
\includegraphics[width=0.47\textwidth]{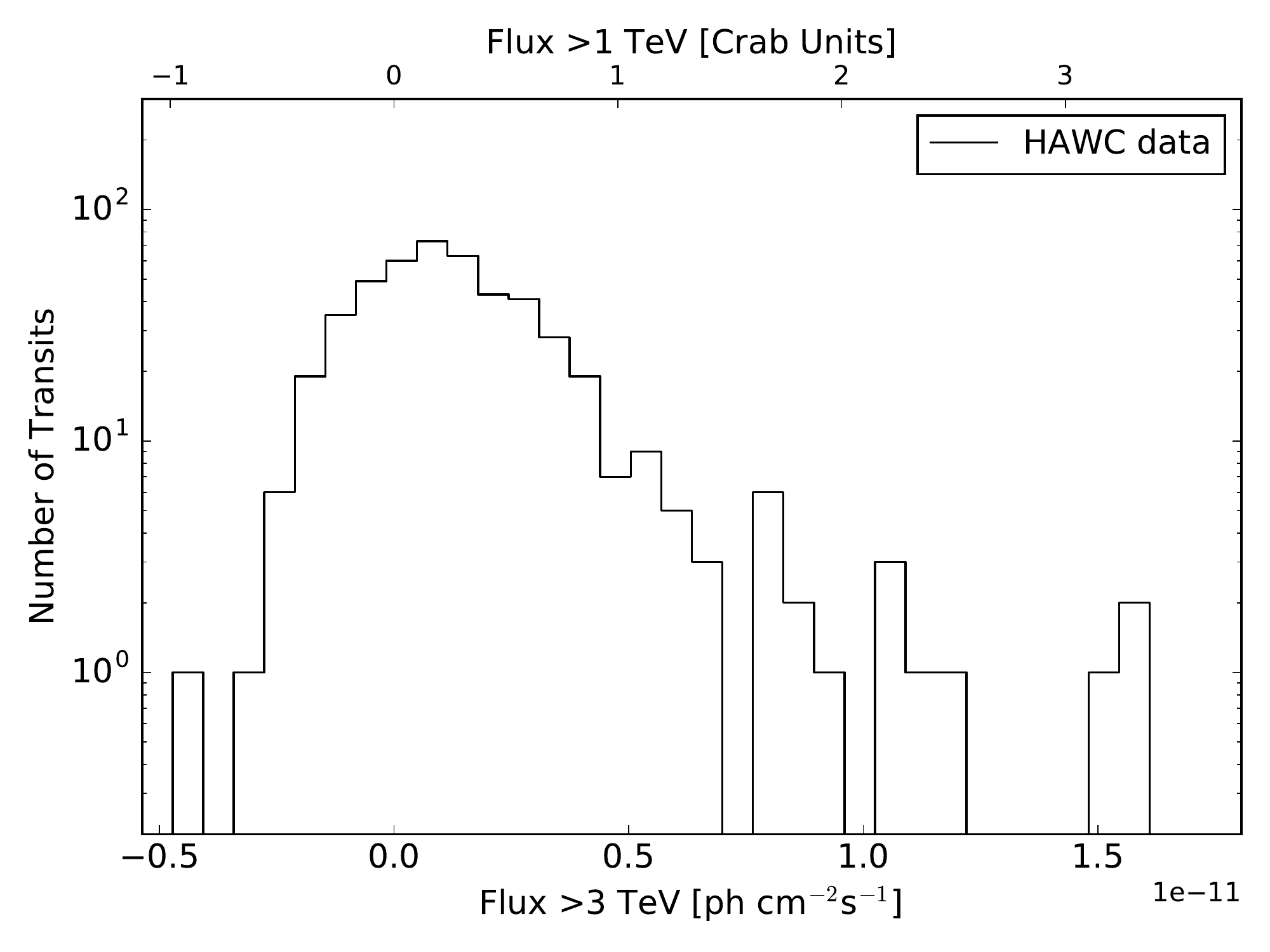}
\caption{Histogram of the 1-transit flux measurements for Mrk 501.}
\label{fig:histFluxMrk501}
\end{figure} 

The TeV light curve for Mrk 501 shows various flaring periods and we find 14 
Bayesian blocks defining distinct flux states under a false detection prior of 
5\%.
The three highest per-transit fluxes exceed the level of three times the Crab 
Nebula flux, corresponding to 0.6\% of all observations included here. 
The last of these flares (2015 August 17) was also captured in tests of the 
HAWC real-time fast transient monitor~\citep{WeisgarberGamma2016} that 
resolves a sub-transit light curve of event rates. 
A histogram of all flux measurements is shown in 
Fig.~\ref{fig:histFluxMrk501}. The shape is generally similar to 
Fig.~\ref{fig:histFluxMrk421}, though with the peak and maximum 
fluxes shifted to lower CU values. As in the case of Mrk 421, the limited 
statistics currently prevent us from distinguishing between different 
functional descriptions of this distribution.

The automated daily light curve monitor that performs the per-transit 
light curve analysis at the HAWC site has been operational since 
2016 February and
identified the increased flux state of Mrk 501 on 
2016 April 6,
is visible on the right side of Fig.~\ref{fig:lcMrk501} at a level 
of $\sim2.4$~CU. We reported this flare in~\citet{ATEL8922}, including the fact 
that the following day still showed a higher than average 
flux before returning to a lower state. The gamma-ray excess rates measured by 
FACT\footnote{See public monitoring 
at \url{http://www.fact­project.org/monitoring} \citep{FACT2013}.} above 
750~GeV also show a rising trend before the HAWC alert.

\subsection{Multiwavelength Comparisons}
\label{sec:mrk501mw}

\begin{figure*}[t]
\centering
\includegraphics[width=0.98\textwidth]{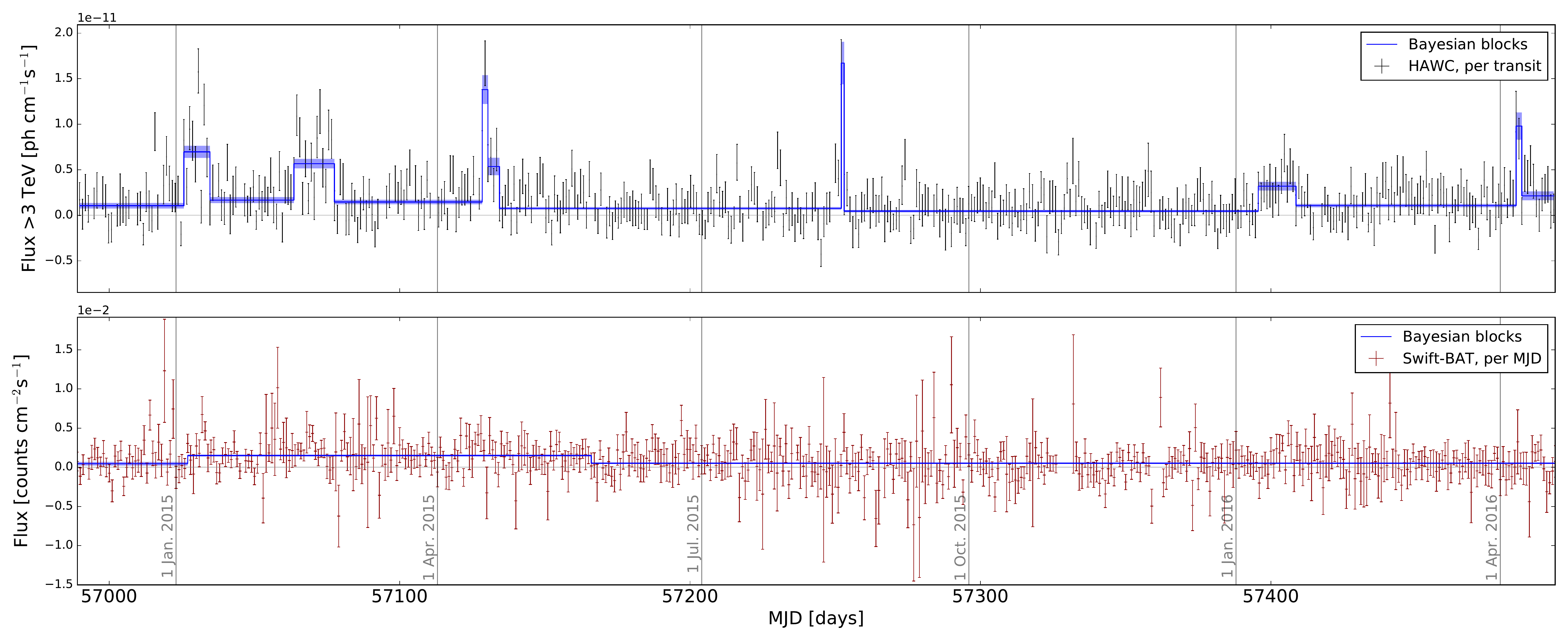}
\caption{Comparison of light curves for Mrk 501 between data from HAWC ($>3$ 
TeV, top panel) and Swift-BAT 
(15 to 50 keV, bottom panel). The results 
of the Bayesian blocks analysis with a 5\% false positive probability are 
included (blue lines).}
\label{fig:Mrk501MW}
\end{figure*}

The Swift-BAT (15 -- 50~keV) data\footnote{Public light curve data from \url{ 
http://swift.gsfc.nasa.gov/results/transients/}.} for Mrk 501 have a very 
high ratio of 
average error 
to weighted mean flux (2.5), but we can apply the same Bayesian blocks analysis 
with prior value $\mathrm{ncp}_{\mathrm{prior}}=6$ as for HAWC data, keeping a 
5\% false 
positive probability. Only 2 change points, i.e. 3 distinct flux states are 
found. The resulting comparison between the three light curves is shown in 
Fig.~\ref{fig:Mrk501MW}.
The X-ray data reveal no 
day-scale light curve features that are correlated with the TeV flares observed 
with HAWC. With the Bayesian blocks 
analysis we find no short flaring periods in the Swift-BAT light curves that 
mirror the activity observed at TeV energies, but the large relative 
uncertainties (average error
corresponds to 2.5 times the mean rate value) precludes us from obtaining a 
quantitative limit for this absence of correlation. 
We can calculate the Spearman rank correlation coefficient for all daily 
averaged fluxes and obtain $0.164 \pm 0.032$, with a 
probability of $10^{-3}$ to occur for an uncorrelated system. This 
positive correlation is qualitatively similar to results obtained previously 
\citep[e.g.][]{Krawczynski2000Mrk501,Tluczykont2010} but is less significant 
than that observed 
for Mrk 421 in Section~\ref{sec:mrk421mw}.

\section{Conclusions and Outlook}
\label{sec:outlook}

We presented the first TeV gamma-ray light curves with sidereal day binning for 
the Crab Nebula, Markarian 421, and Markarian 501 that were obtained with 
the first 17 months of data from the HAWC Observatory. HAWC is currently the 
most sensitive wide-field-of-view TeV gamma-ray observatory and provides unique 
coverage of transients due to its $\sim 95$\% duty-cycle and an unbiased 
daily observation mode.

No variability was found for the Crab Nebula flux measurements, which is in 
agreement with the absence of TeV variability in IACT observations.
For both Mrk 421 and Mrk 501 we found clear variability on time scales of one 
day and use the Bayesian blocks algorithm to identify distinct flux states. 
In the case of Mrk 421, the distribution of unbiased, daily flux measurements 
from HAWC is not well described by a fit to archival TeV data from literature. 
The average flux over the period included here is $\sim 0.8$~CU above 1 TeV, 
significantly higher than previous estimates for an upper limit to the baseline 
flux ($\sim 0.3$~CU) but not exceeding the maximum of past yearly averages. 
The highest flux 
values, averaged over 6 hours, reach up to five 
times the Crab Nebula flux.
Mrk 501, on the other hand, is observed with an average flux $\sim0.3$~CU above 
1 TeV, with flares reaching up to $\sim 3.5$~CU multiple times during our 
observations.
The spectral fit for Mrk 501 is in agreement with previous measurements up 
to a few TeV and shows curvature, modeled 
here as an exponential cut-off at $\sim 6$~TeV.

The public monitoring data for lower energy gamma rays with Fermi-LAT (100~MeV 
to 300~GeV) did not show daily flaring features during the period covered by 
our TeV light curves.
We compared the HAWC data to Swift-BAT X-ray measurements that have similar 
sampling and duty cycle. For daily intervals, we could not identify activity in 
this energy band (15 to 50 keV) that is correlated with the largest TeV 
flaring episodes observed with HAWC. On the other hand, we find positive 
correlations for both Mrk 421 and Mrk 501 between HAWC and Swift-BAT X-ray 
fluxes when 
comparing all daily averaged measurements, similar to previously published 
results. This first look at multiwavelength 
correlations is limited by the low sensitivities of the satellite monitoring 
instruments that result in large uncertainties for average daily 
fluxes. In a forthcoming study, we 
will extend these multiwavelength studies and include data from pointed 
observations with more sensitive instruments where available, in order to 
better assess flux correlations and compare them to broad band model 
predictions.
Ongoing work of improving the energy estimation in the HAWC analysis
will help us to study the spectra of Mrk 421 and Mrk 501 in 
more detail and to investigate changes in spectral behavior 
over time.

The description of the methods, systematic uncertainties, and 
reference applications of the HAWC light curve analysis that we presented here 
provides the basis for day-scale transient studies of any TeV source 
within the approximately two thirds of the sky monitored by HAWC. This analysis 
is already being performed in realtime and will continue to provide flare 
alerts for Mrk 421 and Mrk 501. 
We are in the process of applying this analysis to all candidates listed in  
the HAWC catalog~\citep{HawcCatalog2016}, as well as other target lists, and 
will present those results in a separate publication.

The HAWC Observatory will continue to record unbiased data for every 
source location transiting through its field of view, with an exposure of up to 
6 hours per sidereal day.
With the initial results discussed here and the continuation of this 
analysis program over the following years, we aim to provide 
HAWC TeV light curves as a new resource for studying the time domain of 
astrophysical processes at the highest energies.

\section{Acknowledgements}
We acknowledge the support from: the US National Science Foundation (NSF); the 
US Department of Energy Office of High-Energy Physics; the Laboratory Directed 
Research and Development (LDRD) program of Los Alamos National Laboratory; 
Consejo Nacional de Ciencia y Tecnolog\'{\i}a (CONACyT), M{\'e}xico (grants 
271051, 232656, 260378, 179588, 239762, 254964, 271737, 258865, 243290, 
132197), Laboratorio Nacional HAWC de rayos gamma; L'OREAL Fellowship for 
Women in Science 2014; Red HAWC, M{\'e}xico; DGAPA-UNAM (grants RG100414, 
IN111315, IN111716-3, IA102715, 109916, IA102917); VIEP-BUAP; PIFI 2012, 2013, 
PROFOCIE 2014, 2015;the University of Wisconsin Alumni Research Foundation; the 
Institute of Geophysics, Planetary Physics, and Signatures at Los Alamos 
National Laboratory; Polish Science Centre grant DEC-2014/13/B/ST9/945; 
Coordinaci{\'o}n de la Investigaci{\'o}n Cient\'{\i}fica de la Universidad 
Michoacana. Thanks to Luciano D\'{\i}az and Eduardo Murrieta for technical 
support.

\software{3ML~\citep{Vianello3ML2015}, HEALPix~\citep{HEALPix}, 
ROOT~\citep{ROOT}, NumPy and SciPy~\citep{NumPy}, 
Matplotlib~\citep{Matplotlib}}

\appendix

\section{Bayesian Blocks results}

Tables~\ref{tab:bbTableMrk421} and \ref{tab:bbTableMrk501} show the Bayesian 
block results (5\% false positive probability) from the analysis of 
HAWC daily flux light curves for Mrk 421 and Mrk 501, respectively.

\begin{table}[h]
\centering
\caption{HAWC Bayesian blocks for Mrk 421}
\begin{tabular}{cccc}
\hline
 \hline
MJD Start & MJD Stop  & Duration & Flux $> 2$ TeV  \\ 
          &           &  [days]  & [ph cm$^{-2}$ s$^{-1}$]\\ 
\hline
56988.38 & 56999.64 & 12.01 &  $(1.1 \pm 0.1 ) \cdot 10^{-11}$\\ 
57000.39 & 57005.63 & 5.98 &  $(3.9 \pm 1.3 ) \cdot 10^{-12}$\\ 
57006.37 & 57009.61 & 3.99 &  $(1.6 \pm 0.2 ) \cdot 10^{-11}$\\ 
57010.36 & 57019.59 & 9.97 &  $(6.4 \pm 1.0 ) \cdot 10^{-12}$\\ 
57020.33 & 57020.58 & 1.00 &  $(2.9 \pm 0.4 ) \cdot 10^{-11}$\\ 
57021.33 & 57045.52 & 24.93 &  $(7.2 \pm 0.7 ) \cdot 10^{-12}$\\ 
57046.26 & 57086.40 & 40.89 &  $(3.8 \pm 0.5 ) \cdot 10^{-12}$\\ 
57087.15 & 57090.39 & 3.99 &  $(1.4 \pm 0.2 ) \cdot 10^{-11}$\\ 
57091.14 & 57143.25 & 52.85 &  $(6.9 \pm 0.5 ) \cdot 10^{-12}$\\ 
57144.00 & 57236.99 & 93.74 &  $(4.0 \pm 0.4 ) \cdot 10^{-12}$\\ 
57237.74 & 57239.98 & 2.99 &  $(2.2 \pm 0.2 ) \cdot 10^{-11}$\\ 
57240.73 & 57254.95 & 14.96 &  $(8.1 \pm 0.9 ) \cdot 10^{-12}$\\ 
57255.69 & 57275.89 & 20.94 &  $(1.3 \pm 0.7 ) \cdot 10^{-12}$\\ 
57276.63 & 57319.76 & 43.88 &  $(4.7 \pm 0.5 ) \cdot 10^{-12}$\\ 
57320.51 & 57368.63 & 48.87 &  $(4.8 \pm 4.2 ) \cdot 10^{-13}$\\ 
57369.38 & 57382.59 & 13.96 &  $(5.0 \pm 0.8 ) \cdot 10^{-12}$\\ 
57383.34 & 57387.58 & 5.98 &  $(1.5 \pm 0.2 ) \cdot 10^{-11}$\\ 
57389.32 & 57411.51 & 22.94 &  $(5.3 \pm 0.6 ) \cdot 10^{-12}$\\ 
57412.26 & 57496.28 & 83.77 &  $(1.5 \pm 0.3 ) \cdot 10^{-12}$\\ 
\hline
\end{tabular}
\label{tab:bbTableMrk421}
\end{table}

\begin{table}
\centering
\caption{HAWC Bayesian blocks for Mrk 501}
\begin{tabular}{cccc}
\hline
 \hline
MJD Start & MJD Stop  & Duration & Flux $> 3$ TeV   \\ 
          &           &  [days]  & [ph cm$^{-2}$ s$^{-1}$]\\ 
\hline
56989.66 & 57024.82 & 35.90 &  $(1.1 \pm 0.3 ) \cdot 10^{-12}$\\ 
57025.56 & 57033.79 & 8.98 &  $(7.0 \pm 0.7 ) \cdot 10^{-12}$\\ 
57034.54 & 57062.67 & 28.92 &  $(1.7 \pm 0.3 ) \cdot 10^{-12}$\\ 
57063.46 & 57076.67 & 13.96 &  $(5.7 \pm 0.5 ) \cdot 10^{-12}$\\ 
57077.42 & 57127.53 & 50.86 &  $(1.5 \pm 0.3 ) \cdot 10^{-12}$\\ 
57128.28 & 57129.53 & 1.99 &  $(1.4 \pm 0.2 ) \cdot 10^{-11}$\\ 
57130.28 & 57133.52 & 3.99 &  $(5.4 \pm 1.0 ) \cdot 10^{-12}$\\ 
57134.27 & 57251.20 & 117.68 &  $(7.6 \pm 1.5 ) \cdot 10^{-13}$\\ 
57251.94 & 57252.19 & 1.00 &  $(1.7 \pm 0.2 ) \cdot 10^{-11}$\\ 
57252.94 & 57394.80 & 142.61 &  $(4.5 \pm 1.4 ) \cdot 10^{-13}$\\ 
57395.55 & 57407.77 & 12.96 &  $(3.2 \pm 0.5 ) \cdot 10^{-12}$\\ 
57408.51 & 57483.56 & 75.79 &  $(1.1 \pm 0.2 ) \cdot 10^{-12}$\\ 
57484.31 & 57485.55 & 1.99 &  $(9.8 \pm 1.5 ) \cdot 10^{-12}$\\ 
57486.30 & 57497.52 & 10.97 &  $(2.1 \pm 0.5 ) \cdot 10^{-12}$\\ 
\hline
\end{tabular}
\label{tab:bbTableMrk501}
\end{table}

\pagebreak

\bibliography{lcpaper}

\end{document}